\documentclass[journal]{IEEEtran}
\usepackage{graphics} % for pdf, bitmapped graphics files
\usepackage{epsfig} % for postscript graphics files
\usepackage{times} % assumes new font selection scheme installed
\usepackage{amsmath} % assumes amsmath package installed
\usepackage{amssymb}  % assumes amsmath package installed
\usepackage{subfigure}
\usepackage{url}

\usepackage{pgf,tikz,pgfplots}
\pgfplotsset{compat=1.15}
\usepackage{mathrsfs}

\usepackage{color}
\usepackage{algorithm}

{ % these settings define remarks and example as parts
}

\newcommand{\bi}{\begin{itemize}}
\newcommand{\ei}{\end{itemize}}
\newcommand{\bd}{\begin{displaymath}}
\newcommand{\ed}{\end{displaymath}}
\newcommand{\be}{\begin{eqnarray*}}
\newcommand{\ee}{\end{eqnarray*}}

\makeatletter
\IEEEtriggercmd{\reset@font\normalfont\fontsize{7.9pt}{8.40pt}\selectfont}
\makeatother
\IEEEtriggeratref{1}

\begin{document}
\title{Data-driven Identification of Nonlinear Power System Dynamics Using Output-only Measurements}

\author{Pranav Sharma$^{1}$, \emph{Student Member, IEEE}, Venkataramana Ajjarapu$^{1}$, \emph{Fellow, IEEE} \\ and Umesh Vaidya$^{2}$, \emph{Senior Member, IEEE}
        \thanks{$^1$P. Sharma and V. Ajjarapu are with Department of Electrical and Computer Engineering, Iowa State University, Ames, Iowa, USA. $^2$ U. Vaidya is with Department of Mechanical Engineering, Clemson University, Clemson, South Carolina, USA.
        Financial support from the National Science Foundation's (NSF) and the Department of Energy’s Grants are gratefully acknowledged.}}               

\maketitle
\begin{abstract} In this paper, we propose a novel approach for the data-driven characterization of power system dynamics. The developed method of Extended Subspace Identification (ESI) is suitable for systems with output measurements when all the dynamics states are not observable. It is particularly applicable for power systems dynamic identification using Phasor Measurement Units (PMUs) measurements. As in the case of power systems, it is often expensive or impossible to measure all the internal dynamic states of system components such as generators, controllers and loads. PMU measurements capture voltages, currents, power injection and frequencies, which can be considered as the outputs of system dynamics. The ESI method is suitable for system identification, capturing nonlinear modes, computing participation factor of output measurements in system modes and identifying system parameters such as system inertia. The proposed method is suitable for measurements with a noise similar to realistic system measurements. The developed method addresses some of the known deficiencies of existing data-driven dynamic system characterization methods. The approach is validated for multiple network models and dynamic event scenarios with synthetic PMU measurements. 
\end{abstract}


\section{Introduction}\label{section_intro}
In the age of the data-revolution, the power grid is evolving at an accelerating rate. The abundance of sensor measurements at all levels of operation (such as smart meters, phasor measurement units, weather sensors) is leading towards a more intelligent and aware grid \cite{amin2005toward}. In both academia and industry, efforts are being made to incorporate these sensor measurements in monitoring, operations and control of the power grid. In particular, phasor measurement units (PMUs) are capable of observing voltage and phasors at $30-120 Hz$ frequency, which is suitable to capture electro-mechanical dynamics of power systems \cite{phadke1993synchronized, phadke2008synchronized}. These PMUs observe, record and communicate the voltage and current phasor in real-time to a centralized control center \cite{phadke2008synchronized}.  

In this work, we have developed an Extended Subspace Identification (ESI) framework for nonlinear dynamic characterization for power systems with PMU measurements. In particular, we have demonstrated applications of developed ESI framework for modal identification (eigenvalue tracing), participation factor computation and inertia estimation. We first present a brief discussion on the state-of-the-art for each of these three applications in power systems. 

\subsection{Literature review}
\subsubsection{Prior work in linear and nonlinear modal identification for power systems}
There is a thrust of research in dynamic behavior characterization of the power systems using PMU measurements\cite{de2010synchronized, fan2013extended}. The challenging aspect of data-driven dynamic modeling of power system behavior is its nonlinear nature, noises in the measurements and changing operating conditions.

One such approach of \emph{Prony Analysis} provides a way to estimate the frequencies and damping by decomposing the given signals into characteristic polynomial equations \cite{fan2013extended, hauer1990initial,  hossain2018online, almunif2019pmu}. Prony Analysis is a fast way to identify dominant modes and inter-area modes. It is assumed that the underlying system has linear dynamics. Also, in its basic form the method is susceptible to noises in the measurements. This work is further extended for multiple-channel Prony Analysis \cite{trudnowski1999making} and distributed implementation \cite{fan2017data} for improving the accuracy of estimation. \emph{Matrix Pencil Method} on the other hand construct a Hankel matrix on the dynamic state measurements and decompose the given matrix using Singular Value Decomposition (SVD) \cite{6024892, 1388545}. This approach provides better accuracy as compared to Prony Method in presence of noisy measurements. However, this approach as well assumes linear nature of system dynamics \cite{almunif2020tutorial}. A detailed summary and comparative study of Prony Analysis and Matrix Pencil Method is presented in \cite{almunif2020tutorial}. Another method of \emph{Subspace Identification} has been developed for output-based identification of linear systems \cite{van2012subspace}. This approach is based on the orthogonal relation between the observability matrix and dynamic states for a linear system that is reflected in the system outputs. However, this method is limited in its applications to linear dynamical systems. In the past, attempts have been made to extend this approach for nonlinear systems \cite{harish2005subspace, takeishi2017subspace, runolfsson2018output}. Yet, all these methods consider partial nonlinearity in either the inputs or system dynamics and hence they are not suitable for a general nonlinear system such as power systems.

Power system dynamics are highly nonlinear. A detailed study of nonlinear dynamic system requires much more complex computation as compared to a linear dynamical system. In \cite{thapar1997application, 1525119, liu2006normal, tian2017accurate}, authors proposed an intermediate expression of system dynamics using Normal Form Analysis. In Normal Form Analysis $2^{nd}$ order Taylor series expansion of the original nonlinear system in polynomial form is considered along with the linearized system matrix ($A$). Thus, representing some nonlinearities of the system by approximating $2^{nd}$ (in some cases $3^{rd}$ order \cite{tian2017accurate}) polynomial functions of the given original systems. These studies have shown that such an approximate model is better suited for system dynamic studies under stress and resonance condition. Moreover, Normal Form Method provides better accuracy for trajectory prediction and controller parameter tuning. This Normal Form Analysis is completely model based and is only applicable when complete dynamic algebraic equations (DAEs) of the underlying power system is available. 

\begin{figure}[htp!]
\centering
\includegraphics[scale=.577]{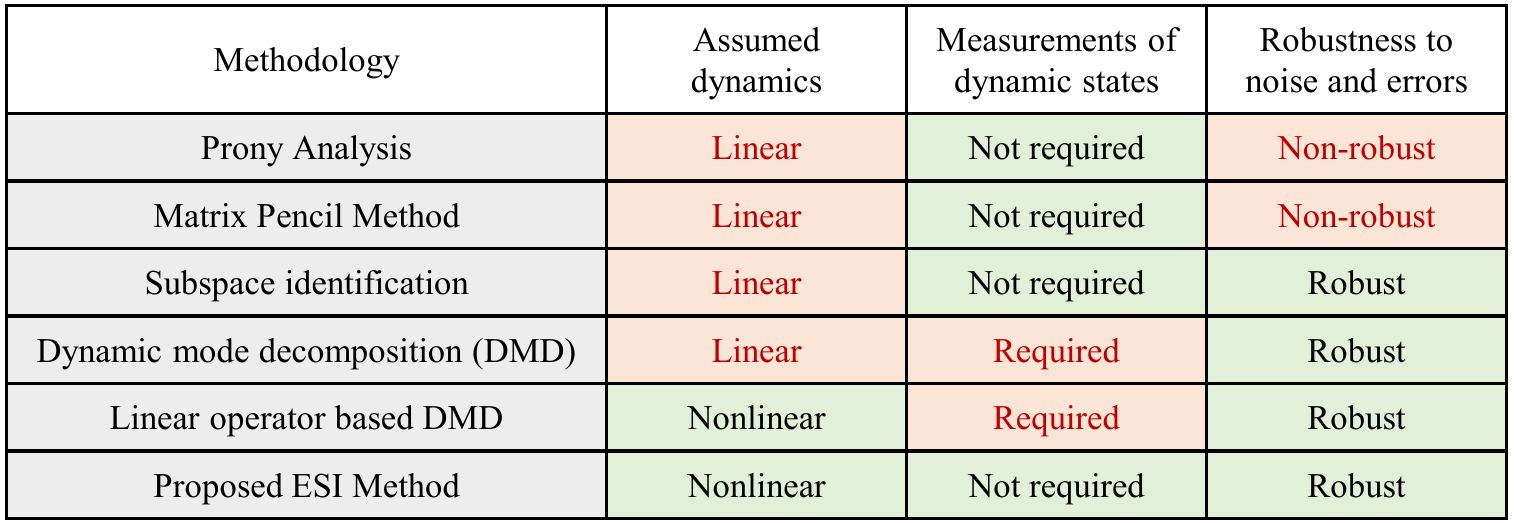}
\caption{Comparison of state-of-the-art data-driven methodologies for power system dynamic characterization }\label{fig_literature_summary}
\end{figure}

One of the recent developments that received great attention from the research community is the linear operator theoretic framework \cite{Koopman1931, korda2018convergence, susuki2014nonlinear, lan2013linearization, tu2013dynamic, susuki2016applied, sinha2018robust, hernandez2018nonlinear}. In particular, the \emph{Koopman operator} provides a linear representation (not the linear approximation) for a nonlinear dynamical system. Recent work has shown that the linear operator based approach is suitable for stability studies\cite{susuki2014nonlinear}, generator coherency identification \cite{5713215}, system inertia estimation \cite{8586007}, and PSS parameter tuning and controller design\cite{korda2018power} and trajectory prediction\cite{8973724}. One of the key assumptions in these works \cite{susuki2014nonlinear}-\cite{8973724} is the availability of state measurements, \textit{i.e.}, it is assumed that time-series measurements are available for all dynamic states of a given system. Though analytically justified, this assumption acts as a roadblock for implementation of these methods for a real power system. In power systems, the dynamic states correspond to internal dynamics of generators, controllers, loads, power electronic converters, etc. For such a system, it is either expensive or impossible to measure all the dynamic states with high fidelity. As in the case of PMUs, we observe voltages, currents and frequencies at the corresponding buses. These measurements are similar to the output of a given dynamical system. In \cite{netto2018robust}, authors have shown that the linear operator framework does not yield meaningful information if they are fed with output measurements such as PMU data.  Fig. \ref{fig_literature_summary} provides a brief summary of state-of-the-art data-driven techniques for power system modal identification. 

\subsubsection{Prior work in data-driven participation factor for power systems}
Originally, the idea of participation factor or selective modal analysis was developed for model based analysis \cite{perez1982selective, verghese1982selective,hashlamoun2009new} of power system dynamics.
This notion of participation factor is very widely used in power system studies for stability analysis, stabilizer design, sensor, controller and renewable placement in the grid. However, existing literature lacks a detailed framework for PMU measurement based participation factor computation or in general data-driven participation factor computation.  
  
  In their recent work \cite{li2020eigensystem}, the authors proposed an eigenvalue realization algorithm based on construction of Hankel matrix from the measurements. Here with given input and output measurements for a completely observable and controllable system, one can construct underlying linear state transition matrix \emph{A, B, C} and thus compute the participation factor. %Firstly, it is considered that the underlying dynamics are linear in nature. 
  In this approach if the system states are not known it is not possible to compute the participation of states in system modes. In \cite{netto_participation_2018}, the authors propose a Koopman operator based framework for nonlinear participation factor computation. However, it is assumed that all the state measurements corresponding to the underlying dynamical system are available. 

\subsubsection{Prior work in system-wide inertia estimation}
In literature, various approaches have been developed for estimation of system inertia using sensor measurements \cite{inoue1997estimation, chassin2005estimation, ashton2014inertia, wall2014simultaneous, tuttelberg2018estimation, su2020adaptive,  8586007}. The central idea of these methods considers that the rate of change of frequency (RoCoF) of a given system is a function of system inertia and difference in input mechanical power and output electric power for a synchronous generator. For $i^{th}$ synchronous generator, this relation can be expressed as: 
  \begin{equation}
      M_i\frac{df_i}{dt} = P_{i M} - P_{i E}  
      \label{eqn_inertia_basic}
  \end{equation}
  Here $\frac{df_i}{dt}$ is the rate of change of frequency at the generator bus, $P_{i M}$ is the input mechanical power, $P_{i E}$ is the output electric power and $M_i$ is the inertia constant for $i^{th}$ synchronous generator. In \cite{ inoue1997estimation, chassin2005estimation}, the authors propose a polynomial approximation of the measurement signal after a transient event to estimate inertia constant. In \cite{ashton2014inertia}, the authors conducted similar studies for Great Britain power system. These methods are appropriate for offline studies of severe events such as loss of generation or load in the system. In \cite{ wall2014simultaneous}, the author presents an inertia estimation and event detection framework using the frequency and power measurements at the sampling rate of 100 measurements/sec. This approach identifies the transient event by applying a smoothing filter to power and frequency measurements and computing system inertia. This approach is online and uses PMU measurements. However, its applications are limited to large transient events. 
  
   A recent work has proven to be capable of estimating inertia from ambient measurements without large transient events \cite{tuttelberg2018estimation}. This approach assumes that an input-output model for various areas in a given power network. The algorithm developed here uses the measurements at the boundaries of respective areas to construct an aggregate dynamic model using autoregressive–moving-average model. This approach is effective in estimating system inertia using ambient data for specific structure of power system model (input-output model) and may not be suitable for a generic system. Furthermore, this requires long window of measurements (5-10 minutes) to estimate the inertia with some accuracy. In \cite{su2020adaptive}, the authors propose a multivariate random forest regression (MRFR) based machine learning approach for inertia estimation from ambient frequency data. This approach is shown to be effective as it reduces the mean absolute \% error in prediction from 12\% in \cite{tuttelberg2018estimation} to 5\%. The key drawback of this framework is its dependency on an elaborate offline training that is required for the given machine learning framework to learn characterization of inertia with respect to various frequency measurements. The effectiveness of the framework is not validated for changing topology and network reconfiguration.% which is commonplace in modern power systems. 
  A Koopman operator based approach is also developed in literature \cite{8586007} for inertia estimation. However, in its current form the approach is only limited to transient event measurements. 
  \subsection{Contributions}
In this work, we propose a novel approach of \emph{extended Subspace Identification} (ESI) for a general nonlinear dynamical system characterization with output-only measurements. The developed approach leverages the functionality of linear operators to capture the underlying nonlinear dynamics and retains the computational framework of Subspace Identification to work with output measurements. The proposed approach assumes stochasticity in system measurements and, therefore, can work well with measurement noises. In particular, the developed approach is suitable for PMU measurements (which act as outputs of dynamical power systems) for characterizing the dynamic behavior of the underlying system. The main contribution of this paper are briefly summarized as follows:

\begin{enumerate}
    \item Developed a robust theoretical framework for output measurement based dynamic characterization for a general nonlinear systems without any information about the underlying system model. 
    \item Formulated a linear and nonlinear modal identification (eigenvalue estimation) technique using PMU measurements for power systems. The developed approach is robust to noises present in the PMU measurements. It can capture Normal Form modes and additional nonlinear modes which are otherwise not captured using existing data-driven techniques. 
    \item Formulated a output based data-driven participation factor that provides a direct map between PMU measurements (sensor nodes) and system modes. This participation factor has practical meaning, as it is computed only using the sensor measurements and can be traced back to the physical components in a power network. Furthermore, this participation factor is capable of computing participation factor for linear and nonlinear modes of the system.
    
    \item Developed a data-driven real-time inertia estimation framework using ambient system measurements. The developed approach doesn't require large training data or transient events for estimating system inertia.  

\color{black}   \end{enumerate}

\textcolor{black}{} The rest of the paper is organized as follows. Section \ref{section_power_system_preliminaries} presents the need for output measurement  based characterization of power system dynamics. In section \ref{section_ESI_theory} we discuss the details of the novel approach that is being proposed in this paper. Section \ref{section_powersystemtestcase} presents the numerical examples for power system dynamic characterization, modal identification, participation factor computation, along with a comparison with existing methodologies. Section \ref{section_conclusion}, concludes the paper with discussion on key contributions and future work.
%%%%%%%%%%%%%%%%%%%%%%%%%%%%%%%%%%%%%%%%%%%%%%%%%%%%%%%%%%%%%%%%%%%%%%%%%%%%%%%%%%%%%%%%%%%%%%%%%%%%%%%%%%%%%%%%%%%%%%%%%%%%%
\color{black} 

\section{Need for output measurements based power systems dynamics characterization}\label{section_power_system_preliminaries}
A power system is a highly nonlinear complex dynamical system. A general power system model can be represented as differential algebraic equations (DAEs): 
\begin{eqnarray}
\dot{x}_{t} = F(x_t) \nonumber\\
0 = G(x_t, y_t) \label{eqn_sys_power_1}
\end{eqnarray}
Here, $x_t \in \mathbb{R}^n$ are the dynamic states of the system and $y_t \in \mathbb{R}^m$ are the output or algebraic variable for the system. Here, $F(x_t)$ represents the dynamics of various system components such as generators, loads, controllers, power electronics devices, etc. For example, in this paper we have considered detailed dynamics for synchronous generators along with exciter and governor dynamics \cite{sauer1998power}. $4^{th}$ order synchronous generator dynamic model can be written as follows: 
 {\small\begin{eqnarray}\label{eqn_generator_dynamic}
\begin{aligned}
&\frac{d\delta_i}{dt}  = \omega_i - \omega_s \\
&\frac{d\omega_i}{dt}  = \frac{T_{m_i}}{M_i} - \frac{E_{q_i}^{\prime} I_{q_i}}{M_i} - \frac{(X_{q_i} - X_{d_i}^{\prime})}{M_i} I_{d_i} I_{q_i} - \frac{D_i (\omega_i-\omega_s)}{M_i} \\
&\frac{d E_{q_i}^{\prime}}{dt}  = -\frac{E_{q_i}^{\prime}}{T_{{do}_i}^{\prime}} - \frac{(X_{d_i} - X_{d_i}^{\prime})}{T_{{do}_i}^{\prime}} I_{d_i} + \frac{E_{{fd}_i}}{T_{{do}_i}^{\prime}} \\
&\frac{dE_{{fd}_i}}{dt}  = -\frac{E_{{fd}_i}}{T_{A_i}} + \frac{K_{A_i}}{T_{A_i}} (V_{{ref}_i} - V_i) 
\end{aligned}
\end{eqnarray}}
The algebraic equations at the stator of the generator are: 
%\textbf{Stator algebraic equations}
{\small\begin{align}\label{eqn_stator_algebraic}
\begin{split}
& V_i \sin(\delta_i - \theta_i) + R_{s_i} I_{d_i} - X_{q_i} I_{q_i}  = 0 \\
& E_{q_i}^{\prime} - V_i \cos(\delta_i - \theta_i) - R_{s_i} I_{q_i} - X_{d_i}^{\prime} I_{d_i}  = 0 \\
& \qquad \text{for}\quad  i = 1,\dots, n_g.
\end{split}
\end{align}}
here, $\delta_i$, $\omega_i, E_{q_i}$, and $E_{{fd}_i}$ are the dynamic states of the generator that correspond to the generator rotor angle, the angular velocity, the quadrature-axis induced emf and the emf of fast acting exciter connected to the generator respectively. The algebraic states $I_{d_i}$ and $I_{q_i}$ are the direct-axis and quadrature-axis currents induced in the generator. $V_i$ and $\theta_i$ are bus voltages and angle. The parameters $T_{m_i}, V_{{ref}_i}, \omega_s, M_i$, and $D_i$ are the mechanical inputs and machine parameters. The stator internal resistance is denoted by $R_{s_i}$ and $X_{q_i}$, $X_{d_i}$, $X_{d_i}^{\prime}$ are the quadrature-axis salient reactance, direct-axis salient reactance and direct-axis transient reactance respectively.

An interesting point to note here is that it might not be possible to measure all the dynamical states of such a system. For example, $ E_{q_i}$, and $E_{{fd}_i}$ are the internal states of the generator and it is not feasible to measure them in real-time. As pointed out in section \ref{section_intro}, majority of the data-driven approaches consider the availability of time-series information for all (or most) state variables ($x(t)$). Existing work on data-driven nonlinear dynamic characterization \cite{susuki2014nonlinear}-\cite{8973724}, either assumes that measurements for these states are available (as they can be generated in simulation studies); or else, a simple swing equation based $2^{nd}$ order dynamics are considered with $\delta_i, \omega_i$ states and thus ignoring other higher order dynamics of system components.
%%%%%%%%%%%%%%%%%%%%%%%%%%%%%%%%
\subsection{PMU measurements as output of power system dynamics}
One of the key advancements in power industry in the past two decades is the deployment of Phasor Measurement Units (PMUs) in electric grid. PMUs are high fidelity, globally synchronized measurement devices that capture voltage and current phasors at a given system node. A typical PMU can capture voltage phasors, current phasors, frequencies and real/reactive power injection for the bus. For this study we have considered PMU measurements at each bus with following measurements ($\vec{V}, \vec{I}, P, Q, f$), namely, voltage phasor, current phasor, real and reactive power and frequency. For example, for the set of equations (\ref{eqn_generator_dynamic}) and (\ref{eqn_stator_algebraic}) mentioned above, we can capture the voltage $\vec{V}_i$, vector sum of $I_{d_i} + I_{q_i}$, active and reactive power ($P_i, Q_i$)  using PMU measurements. %Hence, for the system described by set of equations (\ref{eqn_generator_dynamic}) and (\ref{eqn_stator_algebraic}), 
Therefore, PMU measurements can be considered as the output of the underlying dynamical systems.

\subsection{PMU measurement features}\label{subsection.PMU.measurement.features} PMUs capture the voltage and current phasors at a rate of 30-120 samples per second. At this sampling rate, PMU measurements can observe the electro-mechanical dynamics of the underlying system. For this study, we have assumed a sampling rate of 100 samples/sec. For an actual system, there is bound to be an error and noise in the measurements. As per IEEE standards on PMUs \cite{martin2008exploring}, the upper bound for Signal to Noise Ratio (SNR) is $20dB$ for steady state measurements.
Hence, we have assumed a presence of a random Gaussian noise with $20 dB$ SNR in all measurements.   

\subsection{Assumption on observability} For the given system of equations (\ref{eqn_sys_power_1}), it is assumed that the system is observable with given output measurements $y_t$. We can validate these assumptions for a system of differential algebraic equation (DAE) by constructing observability gramian. However, for a stochastic system with the output measurements, without any information of underlying system of equations, it is not possible to validate full observability. We can estimate the minimum number of perturbed modes using persistency of excitation condition. But that doesn't necessarily translate to the actual number of modes present in the system. 

For this purpose, firstly, we have assumed that PMUs provide a full observability for typical state estimation problem, \textit{i.e.}, we are able to observe all bus voltages and current phasors. Secondly, the PMU sampling is assumed to be 100 samples per second, this provides a reasonable observability for electro-mechanical dynamics and other slower phenomenon such as voltage stability. It is not possible to capture the transient response (of  $<10  ms$ time scale) with PMU measurements and hence they are not in the purview of this research. That is to say, we have made an assumption on observability motivated by the physical characteristics of power system dynamics and PMU measurements. Within the scope of electro-mechanical and other slower dynamical phenomena, this assumption can be justified. 
%%%%%%%%%%%%%%%%%%%%%%%%%%%%%%%%%%%%%%%%%%%%%%
\section{Novel ESI Method for Nonlinear System Identification Using Outputs}\label{section_ESI_theory}
First we will provide a brief overview of the foundational pieces for the developed ESI approach.
%%%%%%%%%%%%%%%%%%%%%%%%%%%%%%%%%%%%%%%%%%%%%%
\subsection{Preliminaries on linear operator theory }\label{subsection_linear_operator_basics}
Consider a nonlinear discrete dynamical system with output measurements described as follows:
\begin{eqnarray}
x_{t+1}&=&F(x_t)\nonumber\\
y_t&=&G(x_t)+v_t \label{eqn_sys_discrete}
\end{eqnarray}
where ${x}\in\mathbb{X}$ is the state, and ${F}:\mathbb{X} \to \mathbb{X}$ is a nonlinear vector-valued map. For such a system we can define an infinite set of observable functions $\Psi:\mathbb{X}\to \mathbb{C}$, acting on system states $x$. For such a system, we can define a linear infinite dimension Koopman operator $\mathcal{K}$, such that \cite{Koopman1931, korda2018convergence, sinha2018robust}:
\begin{equation}\label{eqn_koopman_basic}
\mathcal{K} \Psi(x)  :=  \Psi\circ {F}(x).
\end{equation}

Koopman operator lifts a general nonlinear system into an infinite dimensional nonlinear function space. Hence provides a linear representation and not a linear approximation. 
%%%%%%%%%%%%%%%%%%%%%%%%%%%%%%%%%%%%%%%%%%%%%%
\subsection{Preliminaries on Subspace Identification}
Subspace Identification is a dynamic system identification method that has been developed for linear systems \cite{van2012subspace}. Consider a linear stochastic dynamical system: 
\begin{eqnarray}\label{eqn_stoch}
\begin{aligned}
& x_{t+1} = Ax_t + w_t\\
& y_{t+1} = Cx_t + v_t 
\end{aligned}
\end{eqnarray}
here $x_t \in \mathbb{R}^n$, $y_t \in \mathbb{R}^l$ are system states and output with $w_t \in \mathbb{R}^n$ and $v_t \in \mathbb{R}^l$ as process and measurement noises. For this we have time-series information for output measurements $\emph{Y} = \{ y_0, y_1, \ldots, y_{s-1} \}$. For a system completely observable with given output $\emph{Y}$ with sufficiently large samples, we can construct a Hankel matrix $Y_{0|2i-1}$, such that
\begin{eqnarray}
\label{eqn_hankel}
Y_{0|2i-1} =  \begin{pmatrix}
y_0 & y_1 & \ldots & y_{j-1} \\
%y_1 & y_2 & \ldots & y_j \\
\vdots & \vdots & \ddots & \vdots \\
y_{i-1} & y_i & \ldots & y_{i+j-2} \\
\hline 
y_{i} & y_{i+1} & \ldots & y_{i+j-1} \\
\vdots & \vdots & \ddots & \vdots \\
y_{2i-1} & y_{2i} & \ldots & y_{2i +j -2}
\end{pmatrix} = \frac{Y_{0|i-1}}{Y_{i|2i-1}} = \frac{Y_p}{Y_f}
\end{eqnarray}
here, $i$ is chosen such that $i \geq 2n$, where $n$ is the expected order of the system and $j = s-2i+1$. This $Y_{0|2i-1}$ matrix is further divided into two equal data sets, referred as past measurements ($Y_p$) and future measurements ($Y_f$). We can define  the orthogonal projection ($\mathbb{O}_i$) of $Y_f$ on $Y_p$, as: 
\begin{eqnarray}\label{eqn_orthogonal_projection}
&& \mathbb{O}_i = Y_f/\textbf{Y}_p = Y_f Y_p^T \cdot (Y_p Y_p^T)^{\dagger} \cdot Y_p
\end{eqnarray}
here ($^\dagger$) denotes pseudo inverse of a matrix. Subspace Identification theorem states that, the orthogonal projection ($\mathbb{O}_i$) is equal to the product of observability matrix ($\Gamma_i$) and Kalman filter states ($\hat{X}_i$), i.e.,    
\begin{equation} \label{eqn_linear_proof}
Y_f/\textbf{Y}_p = \mathbb{O}_i = \Gamma_i\cdot \hat{X}_i
\end{equation}
Using Singular Value Decomposition (SVD), we can obtain $\Gamma_i$ and thus Kalman filter states can be written as $\hat{X_i} = \Gamma_i^\dagger\mathbb{O}_i$. As a result, we can solve the least square problem to obtain $\hat{A}$ and $\hat{C}$ matrix for the underlying linear system, as:
\begin{equation*}
\begin{pmatrix}
\hat{A} \\ \hat{C}
\end{pmatrix} = \begin{pmatrix}
\hat{X}_{i+1} \\ Y_{i|i}
\end{pmatrix} . \hat{X_i}^\dagger
\end{equation*}

Interested readers are encouraged to read \cite{van2012subspace} for more detailed discussions on linear Subspace Identification. 
%%%%%%%%%%%%%%%%%%%%%
\subsection{Observer design from output measurements} \label{subsection_observer design}
A stochastic power system behavior can be represented as:   
\begin{eqnarray}
x_{t+1}&=&F(x_t)+w_t=T(x_t,w_t)\nonumber\\
y_t&=&G(x_t)+v_t\label{eqn_power_sys_2}
\end{eqnarray}
where $x_t\in \mathbb{R}^n$ are the states, $y_t\in \mathbb{R}^l$ are the outputs, and $w_t$ and $v_t$ are the process and measurement noises respectively. Using the functionality of linear operators as shown in equation (\ref{eqn_koopman_basic}), we can select a set of basis functions $\Psi(x)=[\psi_1(x),\ldots, \psi_N(x)]$, such that:
{\small\begin{equation}\label{eqn_state_in_span_of func}
\Psi[F(x)]=\sum_{j=1}^N{\bf K}_{j}\psi_j(x)
\end{equation}}
i.e., the action of the Koopman operator on the basis function $\Psi$ is closed. Further, we also assume that the output function  $G\in {\rm span} \{\psi_i\}$ i.e., 
\begin{equation}\label{eqn_output_in_span_of func}
G(x)=\sum_{j=1}`^N c_j \psi_j(x)
\end{equation}

Using the relation obtained in equation (\ref{eqn_state_in_span_of func}) and (\ref{eqn_output_in_span_of func}), we can write the given power system model (\ref{eqn_power_sys_2}) in the lifted function space, as:
{\small\begin{eqnarray}\Psi(x_{t+1})&=&{\bf K}^\top \Psi(x_t)+\delta_t\nonumber\\
y_t&=&C \Psi(x_t)+v_t \label{sys_lift}
\end{eqnarray}}
Here, matrix $C$ is obtained by stacking the column vector $c_j$. Now our objective here is to retrieve an equivalent system just from the output measurements $y_t$. For this, we first lift the output measurements in higher dimensional functional space, such that:
{\small\begin{eqnarray}\label{eqn_y_lifting_relation}
 \mathbb{R}^l\ni y_t\Longrightarrow Y_t:=\Phi(y_t)\in \mathbb{R}^L,\;\;\;L>>l
\end{eqnarray}}
where $\Phi(y)=[\phi_1(y),\ldots,\phi_L(y)]$ are nonlinear functions used in the lifting of the output under the assumption that $\Phi\circ G(x)\in {\rm span }\{\psi_i\}$ i.e., 
{\small\[\Phi\circ G(x)=\sum_j m_j \psi_j(x)\]}
we can write the lifted output as 
$Y_t=M \Psi(x_t)+\xi_t$. Here $M\in \mathbb{R}^{L\times N}$ matrix obtained by staking $m_j$ as column vectors and $\xi_t$ is the noise in the lifted space. Thus, the given power system model can be written in the lifted function space, as: 
{\small\begin{eqnarray}\Psi(x_{t+1})&=&{\bf K}^\top \Psi(x_t)+\delta_t\nonumber\\
Y_t&=&M \Psi(x_t)+\xi_t \label{eqn_sysoutput_lift}
\end{eqnarray}}
here $\Psi(x_t) \in \mathbb{R}^K$, $Y_t \in \mathbb{R}^L$. For the numerical studies conducted for power systems as discussed in section \ref{section_powersystemtestcase}, we have considered $n^{th}$ order polynomial functions for lifting $\Phi(y_t)$, where $n \in \{3, \cdots 8 \}$ depending upon the size of system and accuracy desired. There is an ongoing research on selection and design of observables functions \cite{marcos_observable} for power systems. A detailed study on the selection of observable functions is outside the purview of this work.
%%%%%%%%%%%%%%%%%%%%%%%%%%%%%%%%%%%%%%%%%%%%%%%%%%%%%%%
\subsection{System identification in observer space}
The system obtained in equation (\ref{eqn_sysoutput_lift}) provides a linear representation for the underlying power system with only output measurements. For the given system, the pair $\{\textbf{K}^T,M\}$ is assumed to be observable. The extended observability matrix $\Gamma_i$ is defined as:
{\small\begin{eqnarray*}\label{eqn_observability}
\Gamma_i =  \begin{pmatrix}
M \\
M\textbf{K}^T \\
%M(\textbf{K}^T)^2 \\
\vdots \\
M(\textbf{K}^T)^{i-1}
\end{pmatrix}  \in \mathbb{R}^{Li \times K}
\end{eqnarray*}}
Here $i$ is proportional extension in observability matrix. It is important to note that $\Gamma_i$ will have a rank of $N$ for dynamic system to be observable. Similarly, Reversed extended controllability matrix $\Delta_i^c$ can be defined as:
{\small\begin{equation}\label{eqn_controllability}
\Delta_i^c = 
\begin{pmatrix}
(\textbf{K}^T)^{i-1}G   & \cdots & \textbf{K}^TG & G \\
\end{pmatrix}
 \hspace{0.2 in} \in \mathbb{R}^{K \times Li}
\end{equation}}
Here, $G$ is the Lyapunov equation for the covariance of output and states. $\Delta_i^c$ will also be of rank $N$, as the given noise is able to excite all $N$ modes present in the system. In practice it is impossible to determine whether \textit{all modes} of the underlying power system are excited for the given output. For the given output measurements in the function space, block Hankel matrix $\{Y_f, Y_p \}$ can be derived using the relation given in equation (\ref{eqn_hankel}).
\begin{eqnarray}
\label{eqn_hankel_nonlinear1}
Y_{0|2i-1} = \frac{Y_{0|i-1}}{Y_{i|2i-1}} = \frac{Y_p}{Y_f}
\end{eqnarray}
Given that the system is completely observable and controllable, we can write the product of extended observability matrix and extended controllability matrix, as:
{\small\begin{eqnarray}\label{eqn_control_observer_covariance}
\Theta_{[Y_f, Y_p]} = \Gamma_i \Delta_i^c
\end{eqnarray}
where $(\Theta_{[A,B]})$ is the covariance between two matrices $A$ and $B$, defined as:
\begin{eqnarray}\label{eqn_covar_defn}
&& \Theta_{[A,B]} = \lim_{j \rightarrow \infty}\frac{1}{j}\sum_{i=0}^jA_iB_i^T
\end{eqnarray}}
For the given system we can estimate Forward Kalman filter states for the given output as: 
{\small\begin{eqnarray}
&& \hat{\Psi}(x_{t+1})= \textbf{K}^T\hat{\Psi}(x_t) + {K^f} \hat{e}_t \nonumber \\
&& Y_t = M \hat{\Psi}(x_t) + \hat{e}_t
\end{eqnarray}}
Here expectation factor ($\mathbb{E}$) over Kalman states is defined as:
{\small\begin{eqnarray*}
\mathbb{E}[Y_tY_t^T] = M\mathbb{E}[\Psi(x_t) (\Psi(x_t))^T]M^T + \mathbb{E}[\hat{e}_t\hat{e}_t^T]
\end{eqnarray*}}
This Kalman filter relation can be further reduced as solutions of the forward Riccati equation \cite{bittanti2012riccati}. The Kalman states can be written in terms of extended controlability matrix ($\Delta_i^c$)and covariance of Hankel matrix defined on output measurements ($Y_f, Y_p$).
{\small\begin{equation*}
    \hat{\Psi}({x_t}) = \Delta_t^c \Phi_{[Y_p, Y_p]}^{-1} \begin{pmatrix}
    Y_0,   Y_1,  \cdots,  Y_{t-1}
    \end{pmatrix}^{T}
\end{equation*}}
Further, rearranging Kalman filter states $\hat{\Psi}({x}_t)$ in Hankel form as illustrated in equation (\ref{eqn_hankel}) will give us: 
{\small\begin{eqnarray*}
\hat{\psi}(X_i) = \begin{pmatrix}
\hat{\Psi}({x}_i) & \hat{\Psi}({x}_{i+1}) & \ldots & \hat{\Psi}({x}_{i+j-1})
\end{pmatrix} 
\end{eqnarray*}}
Which leads to 
\begin{equation}\label{eqn_X_hat_def}
   \hat{\psi}(X_i) = \Delta_i^c\Theta_{[Y_p, Y_p]}^{-1}[Y_p]
\end{equation}
Substituting $\Delta_i^c$ from (\ref{eqn_control_observer_covariance}) in (\ref{eqn_X_hat_def}), we have: 
\begin{eqnarray}\label{eqn_yf_yp_gamma}
 \hat{\psi}(X_i) = (\Gamma_i^{\dagger}\Theta_{[Y_f, Y_p]})\Theta_{[Y_p, Y_p]}^{-1}[Y_p] \nonumber \\ 
 \Gamma_i\hat{\psi}(X_i) = \Theta_{[Y_f, Y_p]}\Theta_{[Y_p, Y_p]}^{-1}[Y_p]
\end{eqnarray}
Further by using the relation of orthogonal projection (\ref{eqn_orthogonal_projection}), we can write:
\begin{eqnarray}\label{eqn_proof_statement}
\Gamma_i\hat{\psi}(X_i) = \Theta_{[Y_f, Y_p]}\cdot(\Theta_{[Y_p, Y_p]})^{\dagger}[Y_p] \nonumber \\
\Gamma_i \hat{\psi}(X_i) = Y_f/\textbf{Y}_p = \mathbb{O}_i
\end{eqnarray}
The lifted system described in equation (\ref{eqn_sysoutput_lift}) is controllable and observable in the function space, where $\Psi(x_t) \in \mathbb{R}^K$. Thus, $rank(\Gamma_i) = K$ and $rank(\hat{\psi}(X_i)) = K$. $\mathbb{O}_i$ being product of two $K$ rank matrices will also be of rank $K$. Hence, $rank(\mathbb{O}_i) = K$. Expanding equation (\ref{eqn_proof_statement}), we have:

\begin{eqnarray*}
&& \mathbb{O}_i = \begin{pmatrix}
M,  M\textbf{K}^T \cdots M(\textbf{K}^T)^{i-1}
\end{pmatrix}^{T} \begin{pmatrix}
\hat{\Psi}({x}_i) & \ldots & \hat{\Psi}({x}_{i+j-1})
\end{pmatrix}
\end{eqnarray*}
As both \textit{row} and \textit{column} matrices are of rank $K$, we can make following inferences for $\mathbb{O}_i$:

\begin{itemize}
\item Row space of $\mathbb{O}_i$ $=$ row space $ \hat{\psi}(X_i)$.
    \item Column space of $\mathbb{O}_i$ $=$ column space $\Gamma_i$.
\end{itemize}

Hence, by performing Singular Value Decomposition (SVD) on $\mathbb{O}_i$, we can obtain $\Gamma_i$. For a choice of matrices $W_1$ and $W_2$ such that:
{\small\begin{eqnarray}\label{eqn_W1W2_properties}
&& W_1 \in \mathbb{R}^{Li \times Li}\qquad \ni rank (W_1) = Li \\
&& W_2 \in \mathbb{R}^{j \times j}\qquad \ni rank(Y_p) = rank(Y_p\cdot W_2)  \nonumber
\end{eqnarray}}
we can write,
{\small\begin{eqnarray*}
&& W_1\mathbb{O}_iW_2 = \begin{pmatrix}
U_1 & U_2
\end{pmatrix}\begin{pmatrix}
S_1 & 0 \\ 0 & 0 
\end{pmatrix}\begin{pmatrix}
V_1^T \\ V_2^T
\end{pmatrix} \\
&& W_1\mathbb{O}_iW_2 = U_1S_1V_1^T = W_1\Gamma_i\cdot \hat{\psi}(X_i) W_2
\end{eqnarray*}} 
The constraint on $W_1$ and $W_2$ ensure that the rank of $\mathbb{O}_i$ is preserved for the given SVD. As the $rank(W_1\mathbb{O}_iW_2) =K$, $S_1$ will be a diagonal matrix of size $K$.
{\small\begin{eqnarray}
&& W_1\Gamma_i = U_1\sqrt{S_1} \nonumber \\
&& \Gamma_i = W_1^{-1}U_1\sqrt{S_1}
\end{eqnarray}}
Thus, Kalman filter states can be written as: 
{\small\begin{eqnarray}\label{eqn_compute_PsiX_i}
&&\hat{\psi}(X_i) =  \Gamma_i^{\dagger}\mathbb{O}_i \\
&& \hat{\psi}(X_{i+1}) = \Gamma_{i-1}^{\dagger}\mathbb{O}_{i-1}
\end{eqnarray}}

Using equation (\ref{eqn_compute_PsiX_i}) and (27), we obtain a linear representation of system as: 
{\small\begin{eqnarray}\label{eqn_final_system_matrix}
\begin{pmatrix}
\hat{\psi}(X_{i+1}) \\
Y_t
\end{pmatrix} = \begin{pmatrix}
\Tilde{K} \\
\Tilde{M}
\end{pmatrix} \begin{pmatrix}
\hat{\psi}(X_i)
\end{pmatrix} + \begin{pmatrix}
\rho_{\delta} \\ \rho_{\xi}
\end{pmatrix}
\end{eqnarray}}
 Here all variables $ \begin{pmatrix}
 \hat{\psi}(X_i) & \hat{\psi}(X_{i+1}) & Y_t
 \end{pmatrix}  $ are known, thus we can compute $\Tilde{K}$ and $\Tilde{M}$ using least square formulation. %A step-wise implementation strategy is presented in Algorithm \ref{algo_subspace} in the Appendix section. 
 The obtained equivalent linear representation as shown in equation (\ref{eqn_final_system_matrix}) helps us understand various dynamical properties of underlying power system. We can identify dominant modes, compute participation factor of outputs in system modes and estimate system parameters \cite{8586007}.
  %%%%%%%%%%%%%%%%%%%%%%%%%%%%%%%
\subsection{Participation factor for output measurements}\label{subsection.participation.factor.def}
For the given system matrix $\Tilde{K}$, we can obtain eigenvalues ($\gamma_i$) and left eigenvector ($\mathcal{E}$) using SVD. The original measurements $y_t$ can be written in terms of lifted observables $Y_t$, as shown in equation (\ref{eqn_y_lifting_relation}). Here $\Phi(y)=[ \phi_1(y),\ldots,\phi_{L}(y)]$ also includes the actual output measurements ($y_t$). Hence, $y(t)$ can be written as: 
{\small\begin{eqnarray}
y_t = \mathcal{B}(Y_t) = \mathcal{B} \Tilde{M}  \hat{\psi}(X_i)
\end{eqnarray}}
Where, $\mathcal{B} \in \mathbb{R}^{l\times L}$ is a linear map from lifted observables to actual output measurements. For this system, we can compute the Koopman modes $\Omega = \mathcal{B}\Tilde{M}\mathcal{E}^{-1}$, as illustrated in detail in \cite{netto_participation_2018}. 

Using these Koopman modes and eigenvectors, we can define the participation ($\mathcal{P}_{ij}$) of $i_{th}$ measurement in $j_{th}$ mode as:
\begin{eqnarray}\label{eqn_participation_factor_defn}
\mathcal{P}_{ij} = \mathcal{E}_{ij}\Omega_{ij} 
\end{eqnarray}
with this, we can compute the participation of output measurements (and corresponding components) in the system modes.
%%%%%%%%%%%%%
 %%%%%%%%%%%%%%%%%%%%%%%%%%%%%%%%%%%%%%%%%%%%%%%%%%%%%%%%%%%%%%%%%%%%
  \begin{figure}[t!]
\centering
\definecolor{qqqqff}{rgb}{0,0,1}
\begin{tikzpicture}[line cap=round,line join=round,x=1cm,y=1cm]
\clip(0,-1.4) rectangle (7.2,3.95);
\draw (0.17, 3.71) node[anchor=north west] {\scriptsize SG1};
\draw (0.17,-0.68) node[anchor=north west] {\scriptsize SG2};
\draw (6.38, 3.71) node[anchor=north west] {\scriptsize SG3};
\draw (6.38,-0.68) node[anchor=north west] {\scriptsize SG4};
%\draw (2.28, 3.32) node[anchor=north west] {\scriptsize DFIG};
\draw (0.60, 2.90) node[anchor=north west] {\scriptsize $1$};
\draw (0.60,-0.10) node[anchor=north west] {\scriptsize $2$};
\draw (6.80, 2.90) node[anchor=north west] {\scriptsize $3$};
\draw (6.80,-0.10) node[anchor=north west] {\scriptsize $4$};
\draw (0.60, 1.70) node[anchor=north west] {\scriptsize $5$};
\draw (1.00, 0.80) node[anchor=north west] {\scriptsize $6$};
\draw (6.80, 1.70) node[anchor=north west] {\scriptsize $7$};
\draw (6.00, 0.80) node[anchor=north west] {\scriptsize $8$};
\draw (1.65, 1.75) node[anchor=north west] {\scriptsize $9$};
\draw (5.40, 1.75) node[anchor=north west] {\scriptsize $10$};
%\draw (1.90, 2.90) node[anchor=north west] {\scriptsize $11$};
%\draw [color=qqqqff](3.65,2.75) node[anchor=north west] {\parbox{1.0 cm}{\scriptsize Area 1}};
\draw [color=qqqqff](1.50,2.90) node[anchor=north west] {\parbox{1.0 cm}{\scriptsize Area 1}};
\draw [color=qqqqff](4.90,2.90) node[anchor=north west] {\parbox{1.0 cm}{\scriptsize Area 2}};
\draw [line width=0.8pt] (0.7,2.8)-- (0.3,2.8);
\draw [line width=0.8pt] (0.7,1.6)-- (0.3,1.6);
\draw [line width=0.4pt] (0.5,2.3) circle (0.2cm);
\draw [line width=0.4pt] (0.5,3.5) circle (0.4cm);
\draw [line width=0.4pt] (0.5,2.1) circle (0.2cm);
\draw [line width=0.4pt] (0.5,2.5)-- (0.5,3.1);
\draw [line width=0.4pt] (0.5,1.9)-- (0.5,1.3);
\draw [line width=0.4pt] (0.5,1.3)-- (1.1,1.3);
\draw [line width=0.8pt] (1.1,1.9)-- (1.1,0.7);
\draw [line width=0.4pt] (0.5,1)-- (1.1,1);
\draw [line width=0.4pt] (0.5,1)-- (0.5,0.7);
\draw [line width=0.4pt] (0.5,0.5) circle (0.2cm);
\draw [line width=0.4pt] (0.5,0.3) circle (0.2cm);
\draw [line width=0.4pt] (0.5,0.1)-- (0.5,-0.5);
\draw [line width=0.8pt] (0.3,-0.2)-- (0.7,-0.2);
\draw [line width=0.4pt] (0.5,-0.9) circle (0.4cm);
\draw [line width=0.4pt] (1.1,1)-- (6.1,1);
\draw [line width=0.8pt] (1.7,0.4)-- (1.7,1.6);
\draw [line width=0.4pt] (2,0.1)-- (2,0.7);
\draw [line width=0.4pt] (1.3,0.1)-- (1.5,0.1);
\draw [line width=0.4pt] (1.5,0.1)-- (1.4,-0.1);
\draw [line width=0.4pt] (1.4,-0.1)-- (1.3,0.1);
\draw [line width=0.4pt] (1.4,0.7)-- (1.4,0.1);
\draw [line width=0.4pt] (1.9,0.1)-- (2.1,0.1);
\draw [line width=0.4pt] (1.9,0)-- (2.1,0);
\draw [line width=0.4pt] (2,-0.3)-- (2,0);
\draw [line width=0.4pt] (1.9,-0.3)-- (2.1,-0.3);
\draw [line width=0.4pt] (1.98,-0.38)-- (2.02,-0.38);
\draw [line width=0.4pt] (1.94,-0.34)-- (2.06,-0.34);
\draw [line width=0.4pt] (1.4,0.7)-- (2,0.7);
\draw [line width=0.4pt] (1.7,1.3)-- (5.5,1.3);
\draw [line width=0.8pt] (5.5,1.6)-- (5.5,0.4);
\draw [line width=0.4pt] (5.2,0.7)-- (5.8,0.7);
\draw [line width=0.4pt] (5.8,0.7)-- (5.8,0.1);
\draw [line width=0.4pt] (5.7,0.1)-- (5.9,0.1);
\draw [line width=0.4pt] (5.9,0.1)-- (5.8,-0.1);
\draw [line width=0.4pt] (5.8,-0.1)-- (5.7,0.1);
\draw [line width=0.4pt] (5.2,0.7)-- (5.2,0.1);
\draw [line width=0.4pt] (5.1,0.1)-- (5.3,0.1);
\draw [line width=0.4pt] (5.3,0)-- (5.1,0);
\draw [line width=0.4pt] (5.2,-0.3)-- (5.2,0);
\draw [line width=0.4pt] (5.1,-0.3)-- (5.3,-0.3);
\draw [line width=0.4pt] (5.14,-0.34)-- (5.26,-0.34);
\draw [line width=0.4pt] (5.18,-0.38)-- (5.22,-0.38);
\draw [line width=0.8pt] (6.1,1.6)-- (6.1,0.7);
\draw [line width=0.4pt] (6.1,1.3)-- (6.7,1.3);
\draw [line width=0.4pt] (6.7,1.3)-- (6.7,1.9);
\draw [line width=0.4pt] (6.7,0.7)-- (6.7,1);
\draw [line width=0.4pt] (6.7,1)-- (6.1,1);
\draw [line width=0.4pt] (6.7,0.5) circle (0.2cm);
\draw [line width=0.4pt] (6.7,0.3) circle (0.2cm);
\draw [line width=0.4pt] (6.7,0.1)-- (6.7,-0.5);
\draw [line width=0.8pt] (6.5,-0.2)-- (6.9,-0.2);
\draw [line width=0.4pt] (6.7,-0.9) circle (0.4cm);
\draw [line width=0.8pt] (6.5,1.6)-- (6.9,1.6);
\draw [line width=0.4pt] (6.7,2.1) circle (0.2cm);
\draw [line width=0.4pt] (6.7,2.3) circle (0.2cm);
\draw [line width=0.4pt] (6.7,2.5)-- (6.7,3.1);
\draw [line width=0.8pt] (6.5,2.8)-- (6.9,2.8);
\draw [line width=0.4pt] (6.7,3.5) circle (0.4cm);
\end{tikzpicture}
\vspace{-.3cm}
\caption{One-line diagram of the two-area system}
\label{fig.2area}
\end{figure}
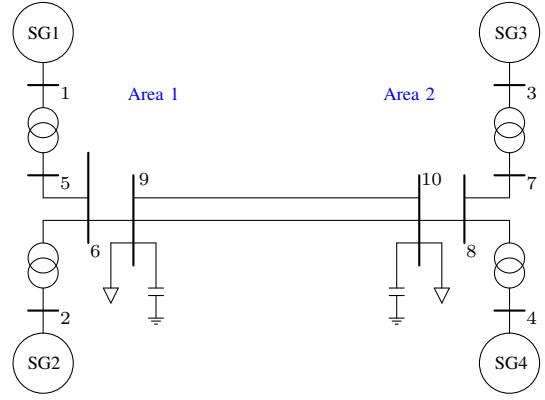
%%%%%%%%%%%%%%%%%%%%%%%%%%%%%%%%%%%%%%%%%%%%%%%%%%%%%%%%%%%%%%%%%%%%%%%%%%%%
\section{Numerical Results}\label{section_powersystemtestcase}
In this section, we will examine the applications of developed ESI approach for power system dynamic characterization. For this purpose, we first consider Kundur 2 area system that is widely used for dynamic studies. Further, results are also shown for IEEE 39 bus system.

\subsection{Results on Kundur 2 Area System}\label{subsection_2_area_results}
As shown in Fig. \ref{fig.2area}, two area (10 bus) system consists of two synchronous generators in each of the two areas with a weak tie line connecting them \cite{kundur1994power}. Here $bus 2$ is considered as the slack bus. %For this system, we have considered a tuned PSS to stabilize system behavior post disturbance.
The system is simulated in a MATLAB simulink environment. Synthetic PMU measurements are recorded for each bus with a sampling rate of $100$ samples/sec. 
\begin{figure}[htp]
\centering
\includegraphics[scale=.55]{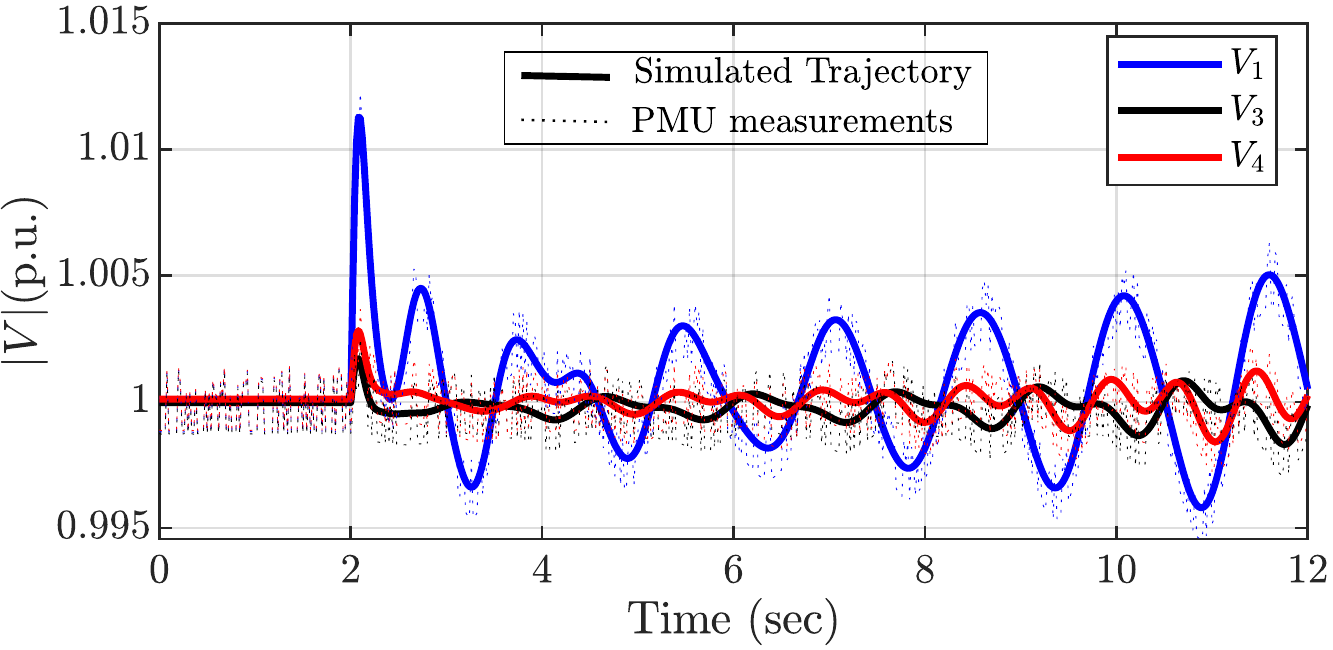}
\caption{Measured bus voltage after perturbation at $t = 2$ sec}\label{Fig.2_area_voltage_plot}
\end{figure}

\begin{figure}[htp]
\centering
{\includegraphics[scale=.53]{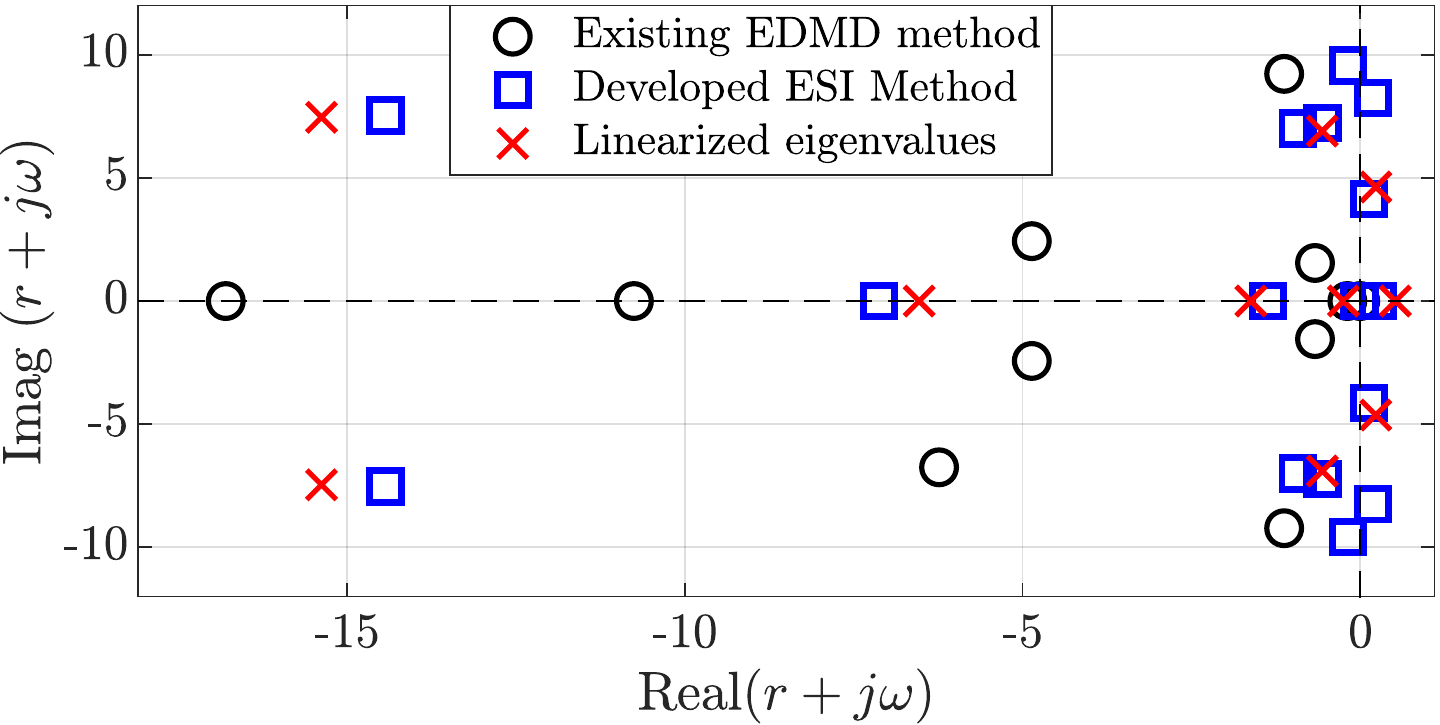}}
\caption{Data-driven eigenvalue estimation from synthetic PMU measurements using existing EDMD method and developed ESI method compared with linearized eigenvalues } \label{Fig.2area_eigv}
\end{figure}

\textbf{1. Modal identification for 2 area system:}
After a small perturbation in voltage at generator bus $SG1$, system undergoes oscillations as seen in the voltage waveform for few selected buses as shown in Fig. \ref{Fig.2_area_voltage_plot}. Here, we considered a weak stabilizer to visualize system oscillations. Moreover, we have considered presence of a Gaussian measurement noise of 20 dB SNR in all synthetic PMU measurements to meet the industry standards. The dotted points shown in Fig. \ref{Fig.2_area_voltage_plot} are considered as the synthetic PMU measurements. As shown in Fig. \ref{Fig.2area_eigv}, ESI method is able to identify linearized system modes present in the system. Here an important point to note is that the extended dynamic mode decomposition (EDMD) method fails to capture the dominant modes accurately with given PMU measurements \cite{susuki2014nonlinear, 8973724}. The same approach works well when state measurements are provided. 

\begin{figure}[t!]
\centering
{\includegraphics[scale=.5]{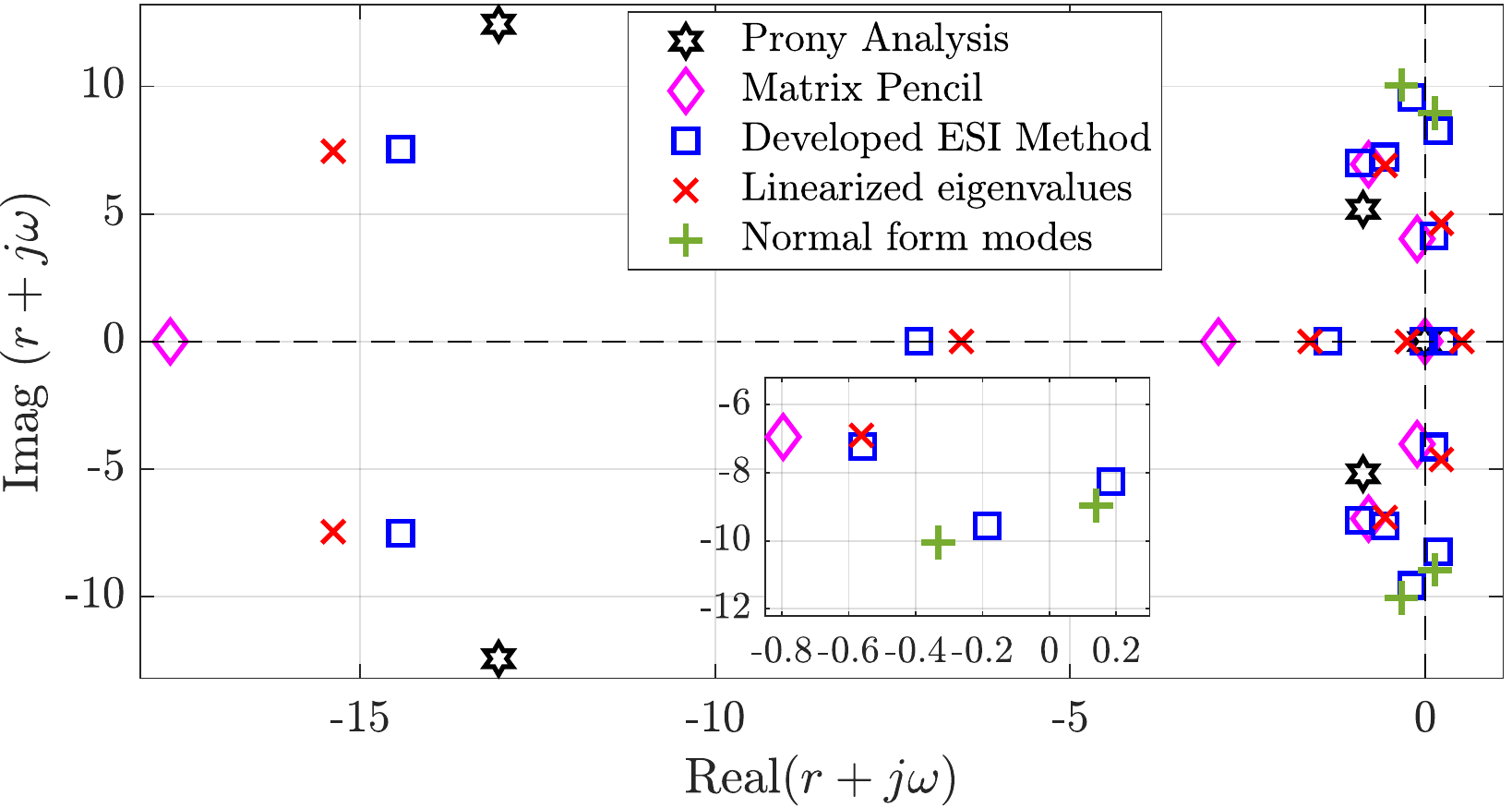}}
\caption{Comparison of data-driven modal identification techinques with linearized and Normal Form modes for the 2 area system } \label{Fig.2area_eigv_nf}
\end{figure}

{We conducted simulation and numerical studies to compare the ESI method with three prevailing techniques, namely, multi-channel Prony Analysis \cite{trudnowski1999making}, Matrix Pencil Method \cite{almunif2020tutorial} and Normal Form Analysis \cite{1525119, liu2006normal}. As shown in Fig. \ref{Fig.2area_eigv_nf}, one can clearly observe that the dominant modes closer to the imaginary ($j\omega$) axis are accurately captured by all techniques. However, the multi channel Prony Method has poor performance compare to the other two.}
 
{We can also see that there are additional modes identified using the ESI approach. This is due to the fact that, we have lifted the given measurements using Koopman operator. In this lifted function space, additional nonlinear modes can be identified. For the given 2 area system, normal modes were computed using the $2^{nd}$ order Taylor series expansion of the given system DAE.}
%%%%%%%%%%%%%%%%%%%%%%%%%%%%%%%%
 \begin{table}[!htbp]%[t!]
\centering \scriptsize
\setlength{\tabcolsep}{0.5em}
\caption{Normal Form modes and corresponding ESI based modes }
\begin{tabular}{c | c | c }
\hline\hline
Domiant Normal & Corresponding Linearized & ESI based \\
form mode & system modes & modal identification \\
\hline
$0.14 + j9.96$ & Mode 2 + Mode 2 & $0.184 + j8.25$ \\
$-0.33 + j10.5$ & Mode 2 + Mode 5 & $-0.18 +j9.56$ \\
\hline\hline
\end{tabular}
\label{tab_nf_mode}
\end{table}
%%%%%%%%%%%%%%%%%%%%%%
 
{Two dominant pairs of Normal Form modes were identified. As shown in Fig. \ref{Fig.2area_eigv_nf}, only the ESI approach was able to capture these nonlinear modes.Table \ref{tab_nf_mode}, provides a comparison of identified Normal Form mode with ESI based mode. We can see that, ESI method is able to capture the Normal Form modes (green +  sign in Fig. \ref{Fig.2area_eigv_nf}). The existing techniques of Prony Analysis and Matrix Pencil Method are not capable of capturing these Normal Form modes of the system.} 

{To summarize the above discussion, ESI method provides an approach for modal identification that is robust to measurement noises. The developed ESI approach has better accuracy as compared to the conventional Prony Analysis and comparable performance with respect to Matrix Pencil Method in identifying linear modes. However, the key advantage of the developed approach lies in its ability to capture Normal Form modes and other nonlinear modes in the higher dimensional Koopman function space.}
%%%%%%%%%%%%%%%%%%%%%%%%%%

\begin{figure}[htp!]
\centering
%\subfigure[]{\includegraphics[scale=.33]{amplitude plot.jpg}}
{\includegraphics[scale=.63]{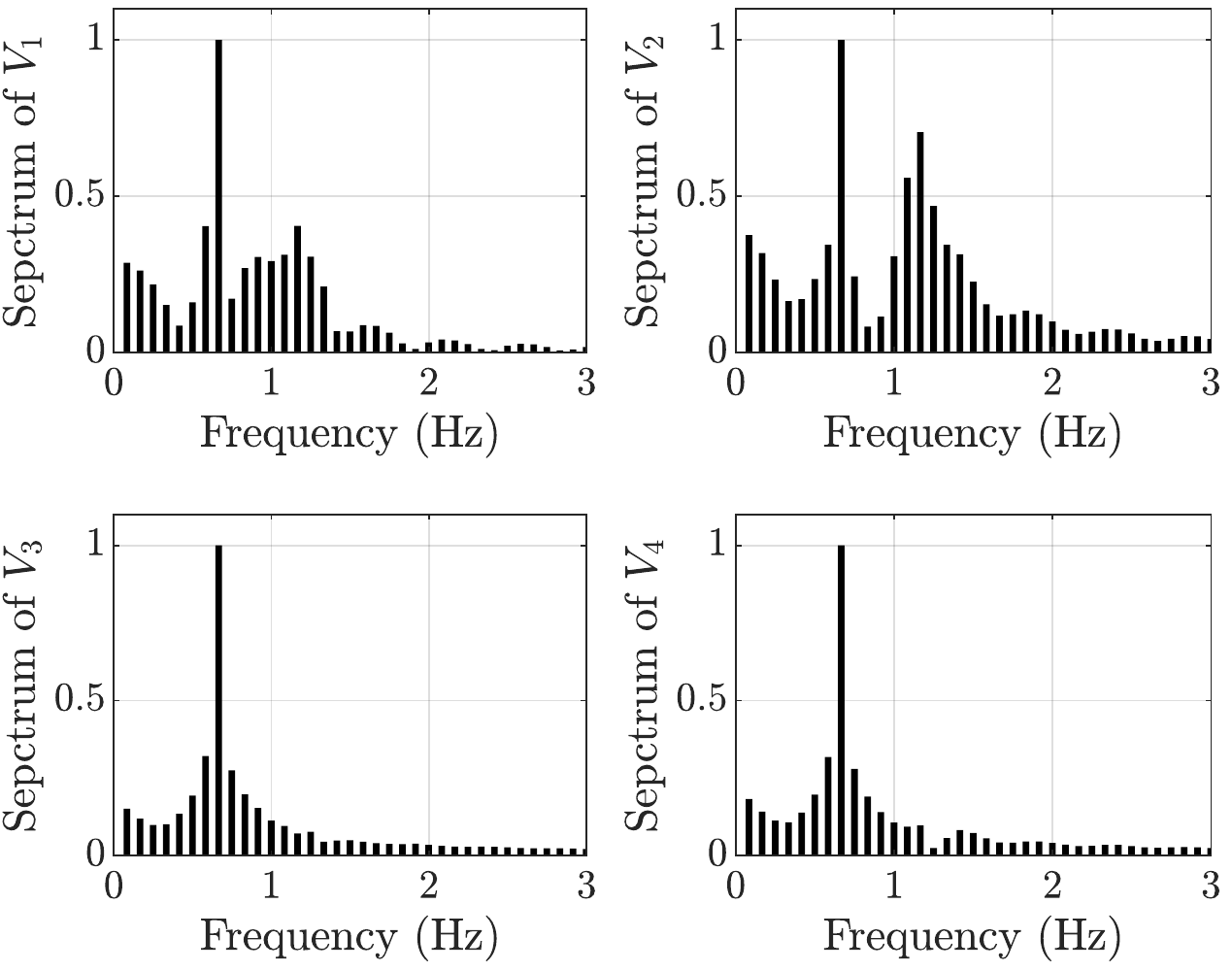}}
\caption{Frequency domain amplitude spectrum plot based on voltage measurements at generator nodes after the disturbance as shown in Fig. \ref{Fig.2_area_voltage_plot}} \label{Fig.2area_spectrum}\end{figure}

\textbf{2. Participation factor of outputs in the dominant modes:}
As shown in section \ref{subsection.participation.factor.def}, we can compute the participation of individual measurements (outputs) in the observed system modes. This helps in mapping the identified oscillations in the system to respective machines and system components. For the 2 area system, three dominant oscillatory modes are identified. For these modes, we have computed participation of measurements corresponding to each of the generator buses. As shown in Table \ref{tab_coherency_2area}, we can locate the mode corresponding to $\approx1.1 Hz$ frequency is a local mode in area 1; similarly, mode corresponding to $\approx0.7 Hz$ frequency is an inter-area mode in which all generators are participating. This inference can be validated by observing signal behavior in the frequency domain. As shown in Fig. \ref{Fig.2area_spectrum}, frequency spectrum plot reveals that $0.7 Hz$ mode is present in all voltage phasors and it is an interarea mode. Similarly, $1.1 Hz$ mode is a local mode observed only in generators corresponding to area 1. These results were also validated using linearized participation factor computation from the given system model. As shown in Table \ref{table_2area_linearized}, participation factor of generator states was computed for related linear modes.
\begin{table}[t!]
\centering \scriptsize
\setlength{\tabcolsep}{0.5em}
\caption{Participation factor of generator bus voltages in dominant Koopman modes for 2 area system}
\begin{tabular}{c c c c c c}
\hline\hline
Frequency & ESI based identified&  \multicolumn{4}{c}{Computed Participation Factor}   \\
(Hz)& modes & $V_1$ & $V_2$& $V_3$& $V_4$  \\
\hline
0.6584 & $0.124 + j4.137$ & 0.346 & 0.275 & 0.168 & 0.211\\
\hline
1.1505 & $-0.559 + j7.229$ & 0.468 & 0.512 & 0.0007 & 0.019\\
\hline
1.1981 & $-14.43 + j7.528$ & 0.492 & 0.506 & 0.000 & 0.002\\
\hline\hline
\end{tabular}
\label{tab_coherency_2area}
\end{table}
%\vspace{-0.5 cm}
%%%%%%%%%%%%%%%%%%%%%%%%%%%%%%%%%%%%%%%%%%%%%%%%%%%%%%%%%%%%%%%%%%%
\begin{table}[t!]
\centering \scriptsize
\setlength{\tabcolsep}{0.5em}
\caption{Model based linearized participation factor for 2 area system}
\begin{tabular}{c c c c c c}
\hline\hline
Linearized eigenvalues & Generator & \multicolumn{4}{c}{Computed Participation Factor} \\
(Frequency) & states & $G_1$ & $G_2$& $G_3$& $G_4$  \\
\hline
$0.23 + j4.63$ & $\delta$ & 0.12 & 0.07 & 0.03 & 0.03 \\
($0.7369 Hz$) & $\omega$ & 0.21 & 0.18 & 0.17 & 0.15 \\
%& $E'_q$ & - & - & - & - \\ 
%& $E_{fd}$ & - & - & - & - \\
\hline
 $-0.562 + j6.9$ &  $\delta$ & 0.14 & 0.08 & - & -  \\
($1.0982 Hz$) & $\omega$ & 0.37 & 0.23 & 0.01 & 0.03  \\
& $E'_q$ & 0.08 & 0.02 & - & - \\ 
& $E_{fd}$ & 0.01 & - & - & -  \\
\hline
$-15.379 + j7.46$ & $\omega$ & 0.02 & 0.02 & - & - \\
($1.1873 Hz$) & $E'_q$ & 0.03 & 0.17 & - & - \\
& $E_{fd}$ & 0.32 & 0.41 & - & -\\
\hline\hline
\end{tabular}
\label{table_2area_linearized}
\end{table}
%%%%%%%%%%%%%%%%%%%%%%%%%%%%%%%%%%%%%%%%
\begin{figure}[htp!]
\centering
\subfigure[]{\includegraphics[scale=.58]{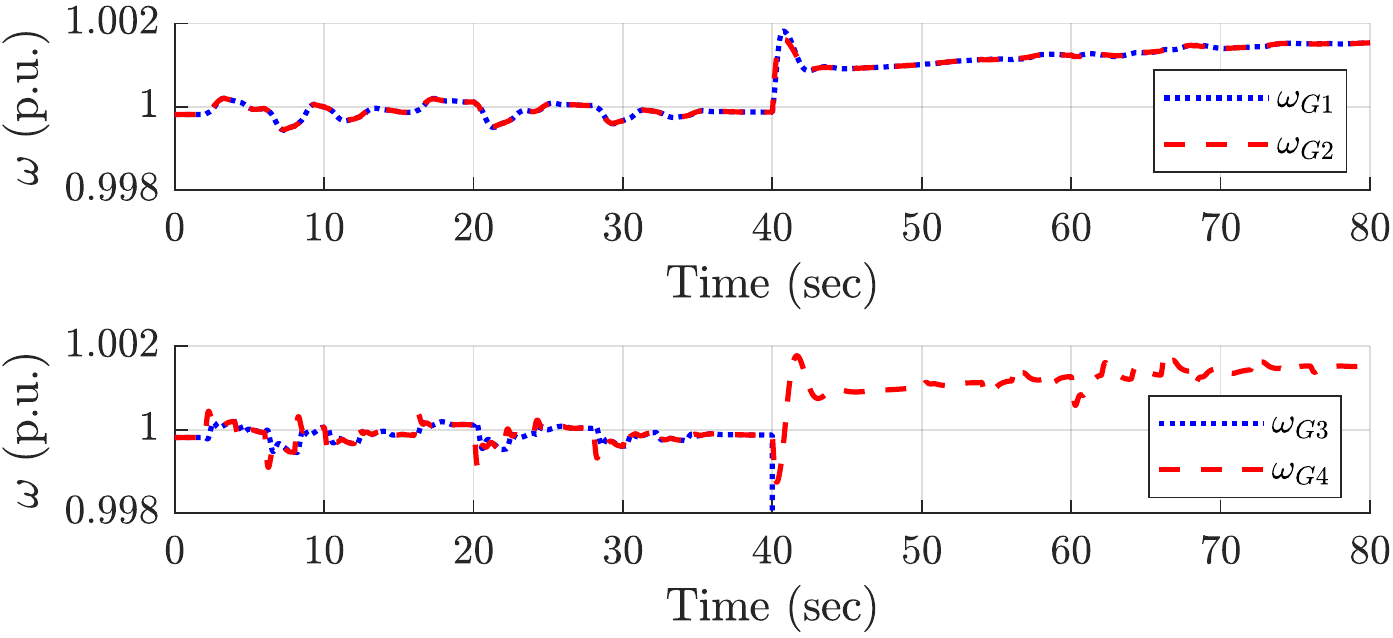}}
\subfigure[]{\includegraphics[scale=.62]{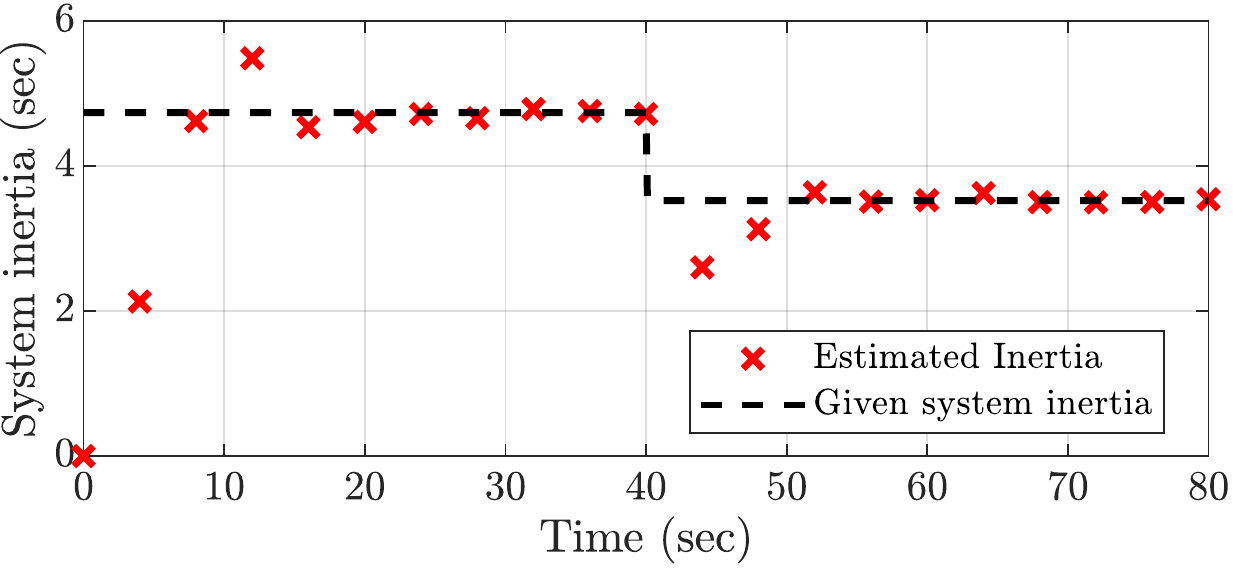}}
\caption{System inertia estimation ($G_3$ is disconnected at t=40 sec) (a) Generator speed response to ambient load change of $\pm 2\%$ (b) System inertia estimation using synthetic PMU measurements}
\label{fig.2_area_inertia_est}\end{figure}

\textbf{3. System inertia estimation:} The developed ESI approach is also suitable for dynamic parameter estimation. As a case study, we have considered the problem of estimating system inertia. For this, we have assumed that PMU measurements are supplemented with generator speed ($\omega_i$) measurements. System parameters such as inertia coefficient can be estimated by projecting the obtained system eigenfucntions on suitable output measurements. This approach is inspired from \cite{5713215}, where authors have used classical model of swing dynamics for synchronous generator and computed system inertia using power injection and rotor speed $\{P_i, \omega_i\}$ measurements. {The central idea is to utilize the relation of power ($P_i$) and rotor speed ($\omega_i$) relation for synchronous generators to estimate the inertia constant ($M_i$) for each generator. 
\begin{equation}
      M_i\frac{df_i}{dt} = \Delta P_{i}  
      \label{eqn_inertia_omega}
  \end{equation}
  Once an equivalent system representation is obtained ($\Tilde{K}$ and $\Tilde{M}$), we simply compute the eigenfunctions corresponding to $P_i, \omega_i$ measurements for each generator. The coefficients of the linear map between the eigenfuctions yields the inertia constant ($M_i$) corresponding to each generator. Here, system inertia $M_{sys}$ is defined as the net sum of inertia constant corresponding to each generator. 
  \begin{equation}
      M_{sys}= \sum_{i=1}^{n} M_i  
      \label{eqn_inertia_total}
  \end{equation}
 Here, $n$ is the total number of generator buses.} As shown in Fig. \ref{fig.2_area_inertia_est} (a), a stable system evolves in presence of ambient disturbances that are emulated by changing the load value (MW). We have considered a 4 sec moving window of measurements to estimate the system inertia. An average of previous and current inertia values is considered to minimize the prediction errors. As shown in Fig. \ref{fig.2_area_inertia_est} (b), we were able to estimate system inertia ($M_{sys}$) accurately (within $5\%$ error) in {first 5 time windows (20 sec)}. At $t=40$ sec, one of the generators ($G_3$) is disconnected from the system and reduces the system inertia. The proposed algorithm is able to estimate the new inertia coefficient {within 4 time windows of measurements.} 
\subsection{Results on 39 bus New England system}
\begin{figure}[htp!]
\centering
\includegraphics[scale=.8]{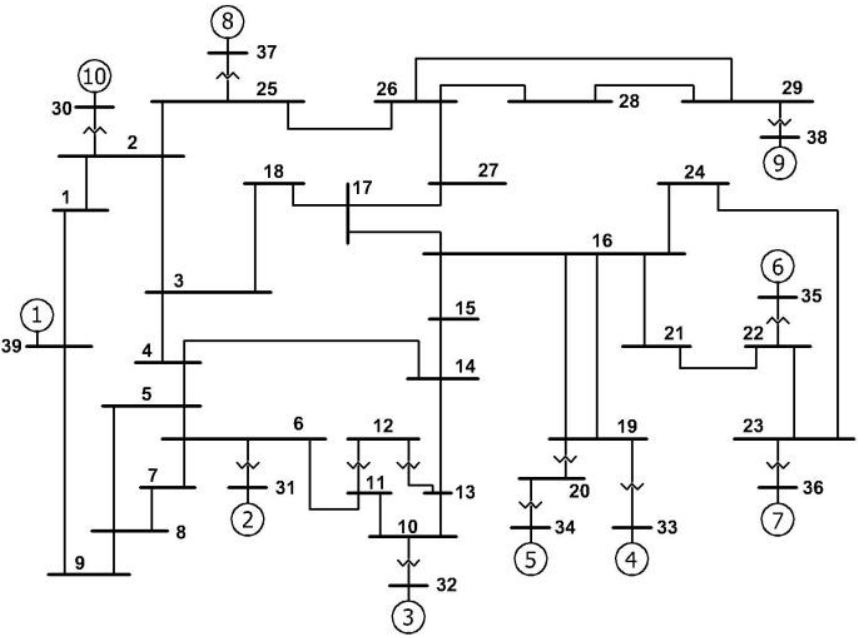}
\caption{One line diagram of the IEEE 39 bus system}\label{Fig.39bus.line.diagram}
\end{figure}

\begin{figure}[htp!]
\centering
\subfigure[]{\includegraphics[scale=.55]{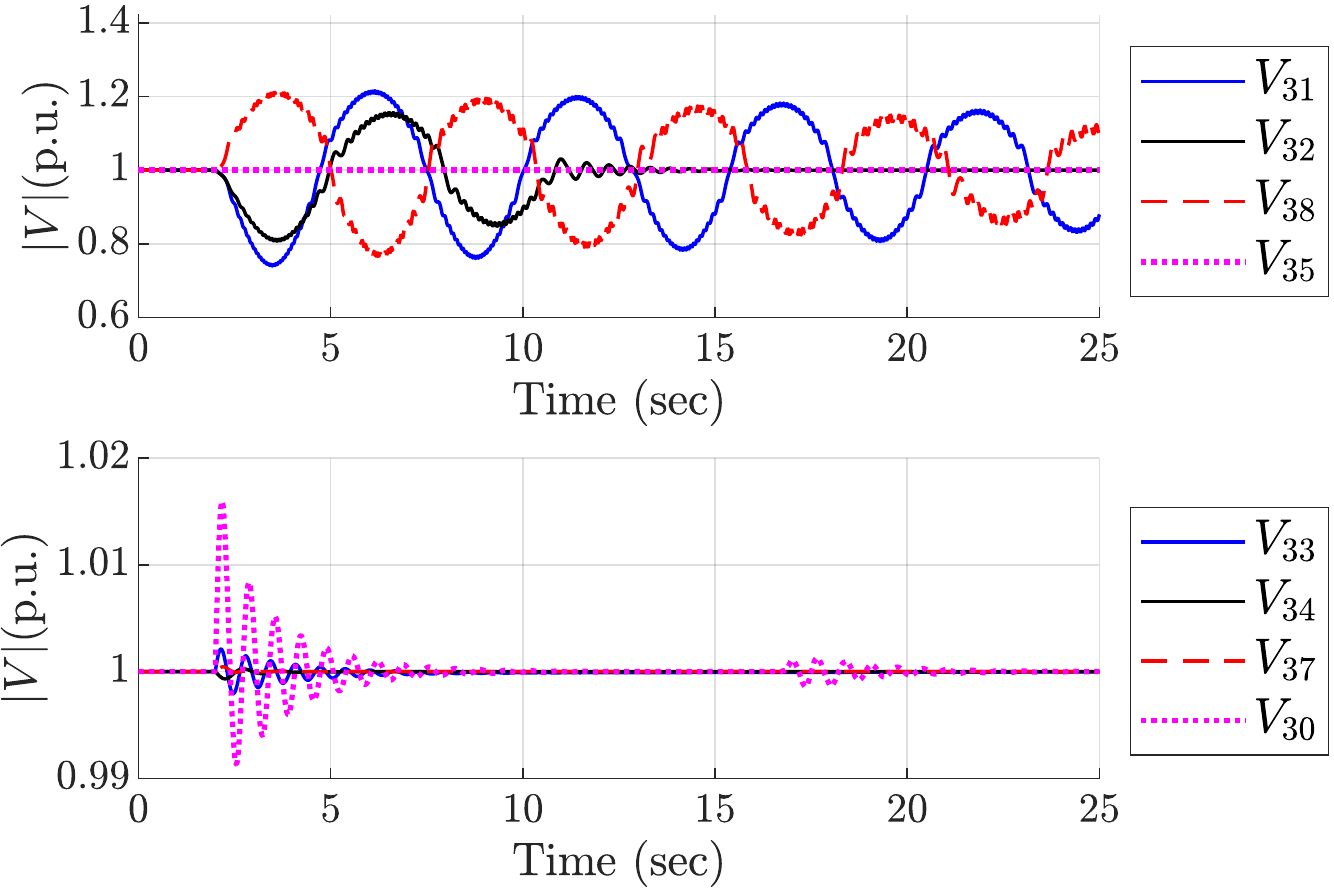}}
\subfigure[]{\includegraphics[scale=.5]{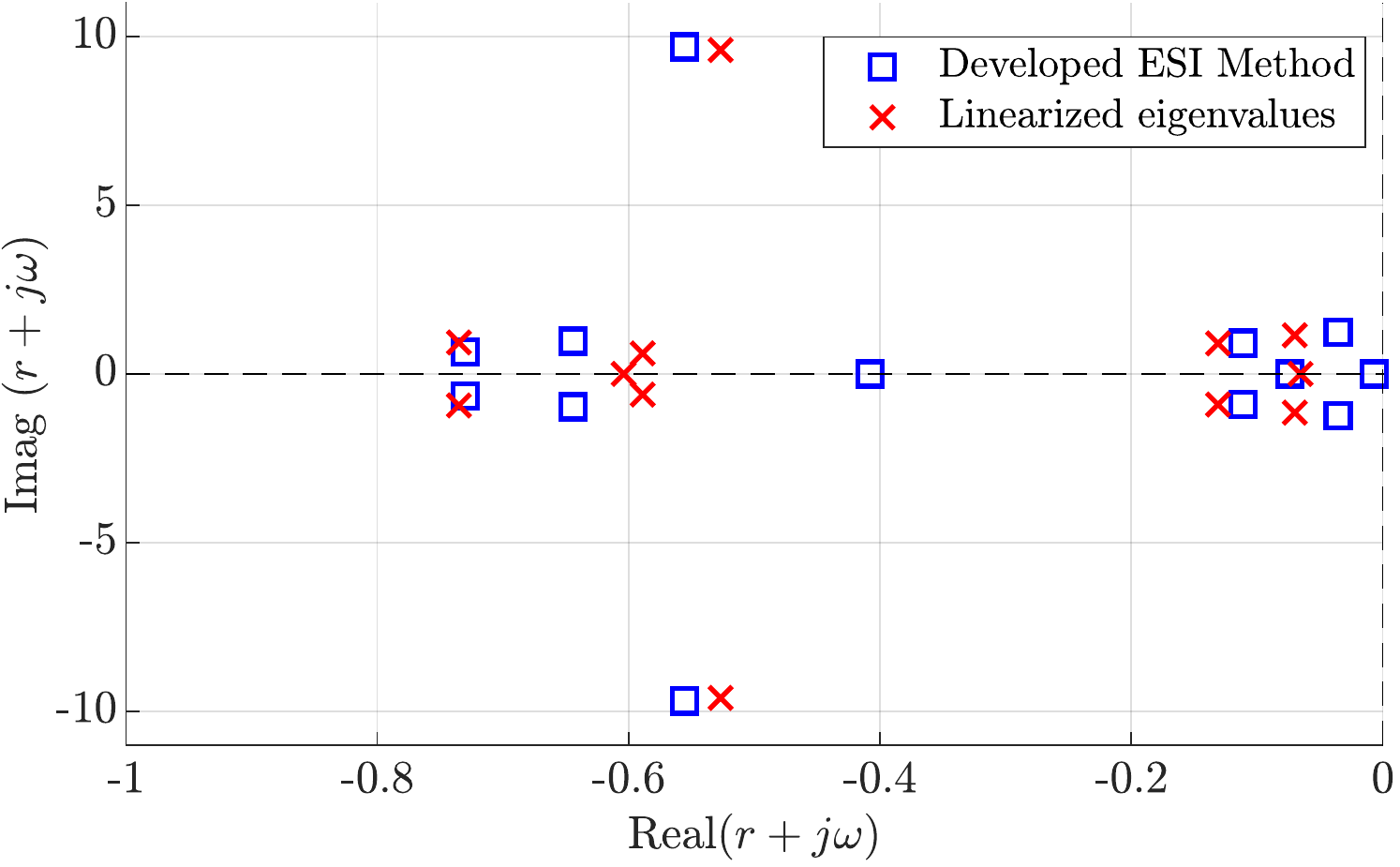}}
\caption{(a) Generator bus voltages in p.u. for 39 bus system after perturbation at $t=2$ sec (b) Estimated eigenvalues using ESI method}\label{Fig.39bus.voltage}
\end{figure}

In the follow-up discussion, we have considered IEEE 39 bus (New England) system to validate the applicability of the developed approach for relatively large system, as shown in Fig. \ref{Fig.39bus.line.diagram}. For the given system, we considered system response after a change in generator set-points  at $t=2$ sec. As shown in Fig. \ref{Fig.39bus.voltage}(a), voltage measurements (as synthetic PMU data) are recorded for all generator and load buses. Using these measurements we were able to identify the linearized system modes, as well as corresponding nonlinear modes, see Fig. \ref{Fig.39bus.voltage}(b). Further, we also computed participation of voltages corresponding to generator nodes in dominant system modes. As shown in Table \ref{table_39bus_participation}, we can map various oscillatory modes present in the system to different generators. 

\begin{figure}[htp!]
\centering
{\includegraphics[scale=.64]{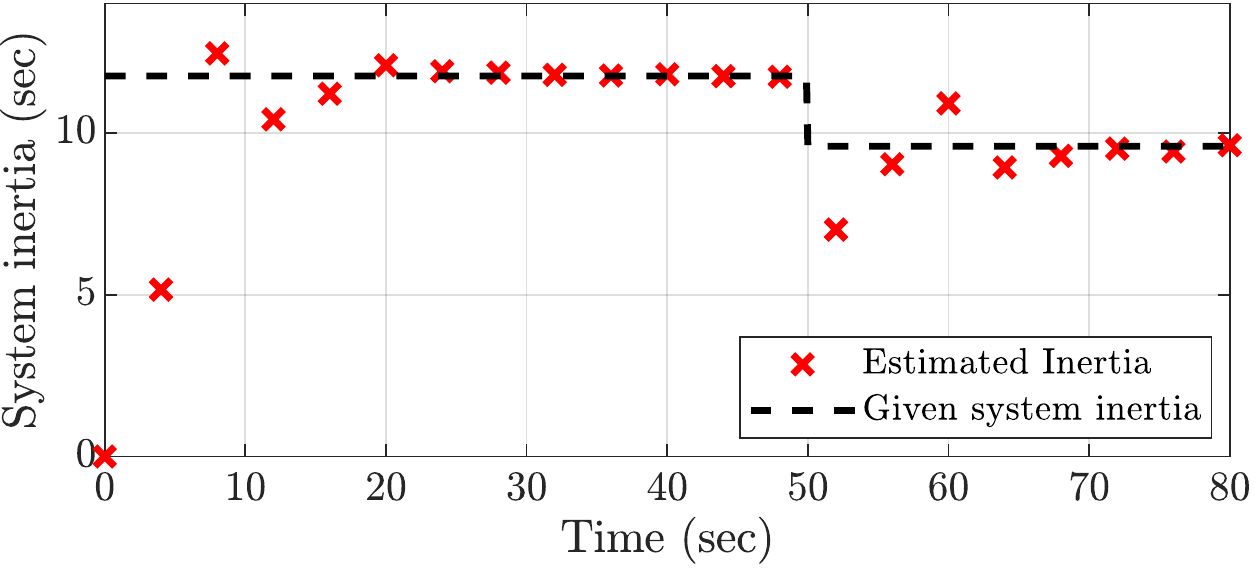}}
\caption{System inertia estimation for 39 bus system ($G_6$ and $G_2$ are disconnected at $t=50$ sec. Inertia estimation with 4 sec moving window}
\label{fig.39bus_inertia_est}\end{figure}

Similarly, the inertia estimation problem is also considered for the 39 bus system. Here, we disconnected generators $\{G_2, G_6\}$ at $t=50$ sec to emulate change in system inertia. Voltage phasors and generator speed ($\omega$) are recorded for ambient load change. The estimation algorithm takes a moving average for predicting overall inertia. As shown in Fig. \ref{fig.39bus_inertia_est}, the ESI method was able to compute system inertia accurately within 4-5 time windows. 

%%%%%%%%%%%%%%%%%%%%%%%%%%%%%%%%%%%%%%%%%%%%%%%%
\begin{table}[t!]
\centering \scriptsize
\setlength{\tabcolsep}{0.5em}
\caption{Output based participation factor for IEEE 39 bus system}
\begin{tabular}{c c c c c}
\hline\hline
ESI based identified & Generator nodes &  Computed  \\
modes & identified & Participation factor \\
\hline
$-0.035 +j 1.238$ &  $V_9$ & 0.481 \\
($0.197 Hz$) & $V_7$ & 0.264 \\
& $V_2$ & 0.209 \\
& $V_{3}$ & 0.167 \\
\hline
$-1.690 + j9.368$ &  $V_7$ & 0.216 \\
($1.488 Hz$) & $V_2$ & 0.186 \\
& $V_{1}$ & 0.104 \\
\hline
$-3.828 + j23.7$ &  $V_{10}$ & 0.421 \\
($3.765 Hz$) & $V_4$ & 0.218 \\
& $V_5$ & 0.095 \\
\hline\hline
\end{tabular}
\label{table_39bus_participation}
\end{table}
%%%%%%%%%%%%%%%%%%%%%%%%%%%%%%%%%%%%%%%%%%%
\section{Conclusion}\label{section_conclusion}

{\subsection{Future Work}
The developed algorithm has computationally intensive steps, namely the Singular Value Decomposition (SVD) and a least square formulation. These steps involve a pseudo inverse computation that has a complexity of $O(n^3)$, where $n$ is the size of matrix under consideration. An immediate extension of this work will be to utilize state-of-the-art efficient algorithms  \cite{dhillon2013new,vega2008fast, kazemi2017fast} for reducing the computation time for least square estimation and pseudo inverse computation to make the developed framework scalable for large systems. The main purpose of this manuscript is to provide an analytical framework and develop its applications for power systems. A discussion on the computation aspects is outside the purview of this paper.} Additionally, the developed ESI framework can also be extended for real-time implementation, feedback control design .  

\subsection{Summary}

Extended Subspace Identification (ESI) method is suitable for nonlinear system identification using output observables. In this paper, we have conceptualized ESI algorithm for output based dynamic characterization of power systems. We have elaborated on the need for output based data-driven techniques for power systems. The algorithm is shown to be robust to noise in sensor measurements within practical limits. 
We have shown applications for modal identification, participation factor computation and parameter estimation (system inertia). Numerical results are shown on a two area system and 39 bus New England system for modal identification with known dynamic properties. A comparative study is also presented with existing data-driven techniques applied to the power systems.

\bibliographystyle{IEEEtran}
\bibliography{main}
\end{document}